# Generation mode-locked square-wave pulse based on reverse saturable absorption effect in graded index multimode fiber

**Zhipeng Dong, Shu jie Li, Jiaqiang Lin, Hongxun Li, Runxia Tao, Chun Gu, Peijun Yao, AND Lixin Xu***

*Department of Optics and Optical Engineering, University of Science and Technology of China, Hefei 230026, China*

*E-mail: xulixin@ustc.edu.cn*

Abstract：We firstly report mode-locked square-wave pulse in Yb-doped fiber laser based on graded index multimode fiber of reverse suturable absorption. By adjusting the pump power, the width of the square-wave can be tuned range from 350 ps to 52.6 ns with 3dB bandwidth spectrum of 0.73 nm. The supermode suppression ratio exceeds ~ 65 dB, which indicates excellent mode-locked operation state, and we also study the characteristic of pulse by the chirp measurement system. The mode-locked square-wave pulse fiber laser can sever as high power light source for industrial applications.

## 1. Introduction

All-fiber-format mode-locked laser has been widely applied among biomedicine, scientific research, and industry due to its compact structure, excellent beam quality and efficient pumping [1]. It is significant to increase the energy of single pulse for the practical application of laser. The traditional soliton pulse energy is limited by the theory of soliton area and is generally limited to the level of hundreds pJ [2]. Diverse approaches have been proposed such as stretched soliton and parabolic pulse, however, these solitons can only output at most a few tens of mJ [3-5]. Akhmediev theoretically investigated dissipative soliton resonance (DSR) pulse forming mechanics based on cubic–quintic Ginzburg–Landau equation (CGLE) [6]. The characteristic of the pulse is that the energy pulse could increase indefinitely as pump power increases while simultaneously the amplitude remains constant, which is represented in the time domain as flat-top or square-wave pulse. This idea attracted a lot of researches, which led to a large number of published research results and improved the pulse energy to the level of μJ [7-9].

To date now, mode-locked square-wave pulse are generally generated based on the nonlinear amplifying loop mirror (NALM) technique [10-12], nonlinear polarization rotation (NPR) mechanism [13]. It is rarely reported that new materials are used as saturable absorbers to generate square-wave pulses due to the damage threshold of these materials is generally low, they cannot withstand large energy pulses. Recently, the nonlinear switching properties or saturable absorber (SA) based on nonlinear multimode interference in graded-index multimode fiber (GIMF) has been widely investigated, which is superiority of low-cost, simple structure, and especially the ability in supporting extremely high damage threshold in high power double cladding fiber laser. It has already used as a saturable absorber in ultrashort fiber laser [14-16], however, there is no mode-locked square-wave pulse by adopting GIMF reported so far.

In this paper, we firstly experimentally demonstrate the reverse saturable absorption effect in GIMF, which adopt our group proposed a new design of saturable absorber based on offset-spliced GIMF, and use it to output mode-locked square-wave pulse in Yb-doped fiber laser. In our fiber laser, the pulse duration can be tuned from 350 ps to 52.6 ns through adjusting the pump power. The maximum energy of output single pulse is 0.44 μJ with a 1.83 MHz repetition rate when pump power reaches 3.24 W. Besides, we also measured the chirp characteristics of square-wave pulse based on the measurement system designed by our group [17].

## 2. Experiment setups

The experiment layout of the mode-locked square-wave laser is shown in Fig. 1. A 3 m double cladding Yb-doped fiber (LIEKKI, Yb1200, 10/125) which pumped by a 915 nm 10 W multimode semiconductor laser through a 2×1 multimode fiber pump combiner. A polarization independent isolator ensures unidirectional propagation, and a polarization controller (PC) is used to adjust optimal mode-locked state. The SA double cladding s fiber (DCF)-GIMF-GIMF-DCF device is placed after the PC. A bandpass filter with a 3 dB bandwidth of 3 nm is utilized to control the central wavelength of the laser. The 40% total energy of laser is extracted by a 60:40 optical coupler and a 99:1 coupler is connected to measure the power, spectral and time-domain characteristics, respectively. The total length of ring cavity is about ~109 m. The instruments used in this experiment are optical spectrum analyzer (ANDO AQ6317B), oscilloscope (Lecroy 640ZI), power meter (Thorlab PM100D), RF spectrum analyzer (AV4021).

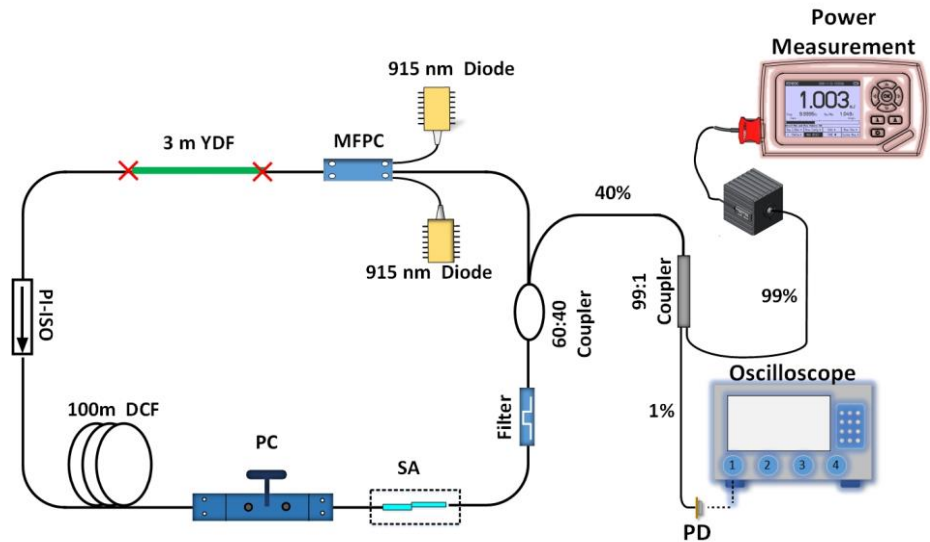

Fig. 1. Experiment schematic of the fiber laser. MFPC, multimode fiber pump combiner; YDF, double cladding ytterbium-doped fiber; PI-ISO, polarization independent isolator; DCF, double cladding fiber; PC, polarization controller; SA, saturable absorber; Filter, bandpass filter;

In order to achieve mode-locked state, the SA device must be fixed in an appropriate curved bending shape. The stable mode-locked square-pulse was generated when pump power reached 569 mW by adjusting the polarization controller. Figure. 2 displayed the characteristic of the square-wave pulse. As shown in Fig. 2(a), the spectrum of square-pulse located at 1063.42 nm with 3dB bandwidth of 0.73 nm. When the pump power increased from 0.57 mW to 3.24 W, the 3dB spectral width decreased from 0.62nm to 0.46nm and the overall shape of spectral remained unchanged. The pulse-duration was tuned from 350 ps to 52.6 ns as the pump power increased, while the pulse amplitude almost remained constant as illustrated in Fig. 2(b), which was consistent with the power clamp theory [6]. The supermode suppression ratio was measured by a RF signal analyzer as shown in Fig. 2(c). The fundamental repetition rate was located in 1.83 MHz with supermode suppression ratio exceeding ~ 65 dB, which indicated excellent mode-locked operation state. The characteristic amplitude envelope modulation is another intrinsic feature of mode-locked square-pulse, which could be observed on the RF spectra gathered for a wider span. The period of modulation was directly connected with the mode-locked pulse-width, and varied continuously with increasing pump power. Figure 2 (d) showed the duration of square-pulse and the variation of the output single pulse energy of pulse with the pump power. When the pumping power reached 3.24 W, the maximum single-pulse energy output of the laser was 139.1 nJ. When the pump power exceeded 3.24 W, the square-pulse would be split and operated in the harmonic mode-locked state. We removed the SA to ensure that the mode-locked mechanism of the laser was supported by the SA. At this time, no matter how we adjusted the pump power and PC, no mode-locked pulse was observed.

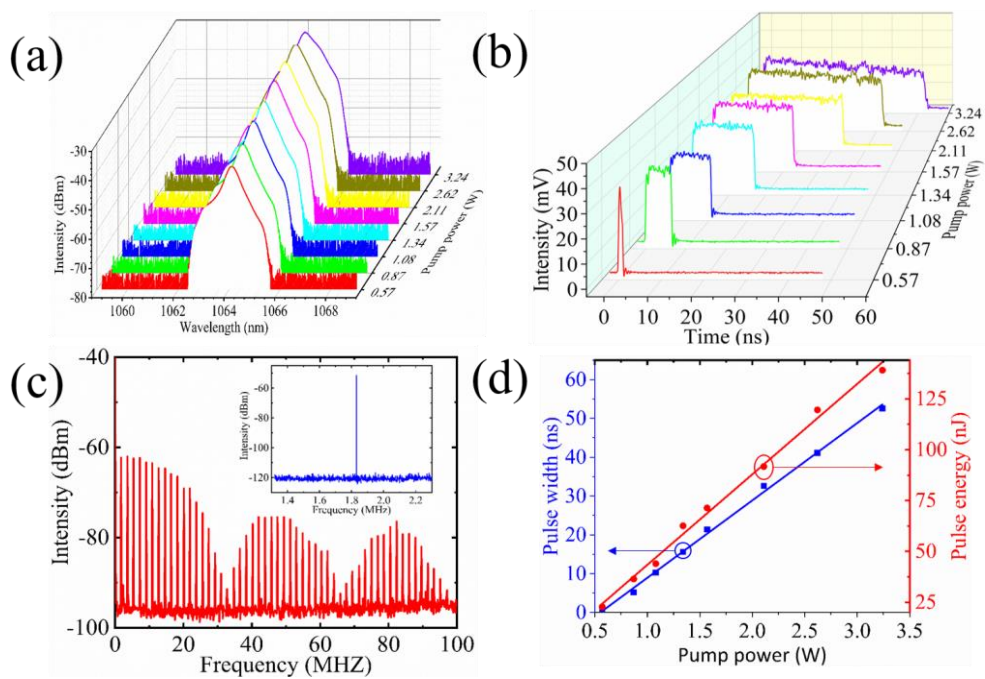

Fig. 2 (a) Output optical spectrum versus pump power; (b) Pulse-duration under different pump powers; (c) RF spectrum of pulse train in the range of 0-100 MHz, inset: RF spectrum with the span of 1MHz and a resolution bandwidth of 10 Hz; (d) Variation of pulse energy and Pulse width versus pump power.

To study the square-pulse chirp characteristic of our mode-locked system, we used the chirp measure device designed by our group, as shown in Fig. 3(a) [17]. In our chirp measure system, it contains two fiber Bragg gratings (FBG1 and FBG2), a polarization independent isolator, a section of 10/125 double cladding fiber of 25 m length, and three 3 dB couplers. Considering that the spectrum of pulse was ranged from 1062.2 nm to 1065.8 nm, the central wavelength of FBG1 and FBG2 was designed 1064.2 nm and 1062 nm with a reflection bandwidth of 0.15 nm and a 99.9% reflectivity, respectively. The pulse of laser was divided into two beams through a 3dB coupler, Pulse from output port A was connected to OSA and oscilloscope for spectrum and waveform measurement. The reflected pulses from port B spectral sampled by grating were also connected to OSA and oscilloscope respectively. The central wavelength of the FBG1 and FBG2 was adjusted by tension to cover the whole spectrum of the pulse and recording the variation of time domain and spectrum of port A, the chirp characteristics of the pulse could be measured. As can be seen from Fig. 3(b), the waveform of the pulse reflected by FBGs versus reflection wavelength. Since the variation of waveform from port B did not change significantly with the change of FGBs reflection spectrum, we consider it to be a randomly chirped pulse.

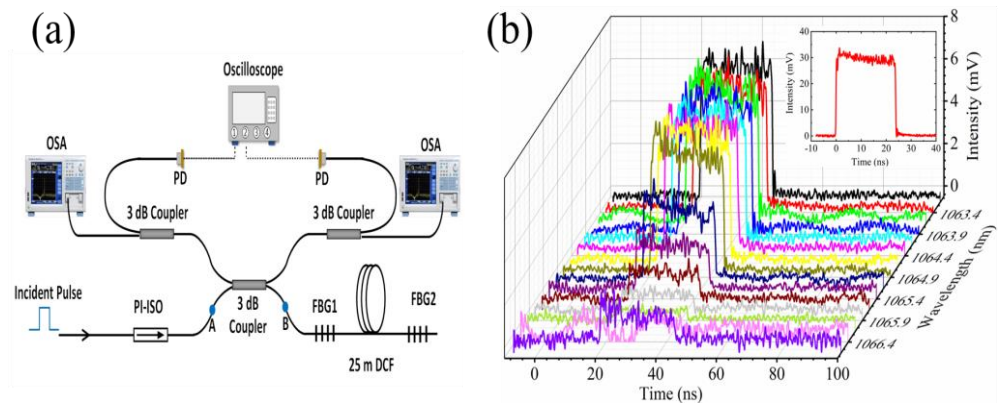

Fig 3. (a)Experimental schematic for pulse chirp system. PI-ISO, polarization independent isolator; FBG, fiber Bragg grating; DCF, corning 10/125 double cladding fiber; OSA, optical spectrum analyzer; (b) Variation of the waveform reflected by FBGs versus reflection wavelength, inset: waveform of pulse from port A;

## 3. Simulation results and analysis

The energy pulse could increase indefinitely as pump power increases while simultaneously the amplitude remains constant, which has already been modeled by cubic-quintic Ginzburg-Landau equation (CGLE), which has additional cubic and quintic saturable absorption terms and quintic nonlinear term, compared with conventional Ginzburg-Landau equation. It should be noted that the quintic saturable term represents the reverse saturation absorption effect, which plays an important role in the formation of pulses from bell to flat top or square [18]. Among the square-wave pulse mode-locked fiber lasers reported in the past, the most common one is based on NPR and non-linear loop mirror mechanism, because they all have sinusoidal nonlinear transmission curve. However, it is rarely reported that the reverse saturable absorption effect is observed in other popular new materials (eg black phosphorus, CNT, WS2, Graphene), because the damage threshold of these materials is relatively low, and the power threshold of reverse saturation absorption effect is generally higher than their damage threshold. The transmissivity decreases with the increase of instantaneous power in a certain range, which is the reverse saturated absorption effect. DSR phenomenon occurs when peak power enters reverse saturated absorption region. Based on this, we believe that there must be a reverse saturable absorption effect in our laser system, and it is formed by GIMF SA rather than NPR mode-locked or nonlinear loop mirror mechanism. Figures 4(a) shows a structural diagram of SA in our laser system. According to the model 8described by Mafi et al [19], and based on the principle of multimode interference in GIMF, the transmittance curve has the characteristics of sinusoidal nonlinear transmission curve. The saturated absorption characteristics of GIMF devices are modeled using the following functions based on our laser system.

$$T = q_1 \sin^2(I/I_{sat}) + q_0$$

Where $q_0 = 0.2$ is linear transmittance, $q_1 = 0.12$ is the modulation depth of device, $I$ is instantaneous power, $I_{sat} = 1\ nJ$ is saturation energy. The numerical simulation results of the transmission curve of the GIMF SA are described in the Fig. 4(b).

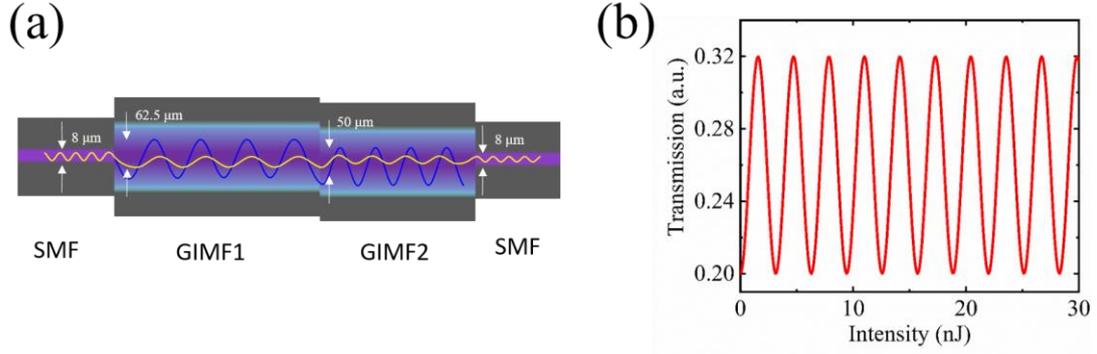

Fig. 4(a) Schematic diagram of the saturable absorber (SA); (b) Numerical simulated of transmission of the GIMF saturable absorber.

### 4. Conclusion

In summary, we have demonstrated a square-wave mode-locked pulse fiber laser based on GIMF SA. We emphasize the importance of reverse saturated absorption in mode-locked square-wave pulse formation, and prove that reverse saturated absorption is caused by GIMF fibers rather than NPR effect or nonlinear optical loop mirror. The pulse width can be tuned form 350 ps to 62.6 ns as pump power increases, which corresponds to single pulse energy from 22.78nJ to 139.06nJ with fundamental repetition rate of 1.83 MHz. In addition, we utilize chirp measurement system to study the chirp characteristic of square-wave pulse, and obtain that the type chirp characteristic of pulse is random chirp. The pulse mode-locked pulse based on GIMF fiber can be used in high power laser to serve in industrial processing.

### Acknowledgments

The authors would thank the National Natural Science Foundation of China (NSFC) (61675188) and Fundamental Research Funds for the central Universities (WK6030000086).